\documentclass[prb,reprint,twocolumn,showpacs,preprintnumbers,amsmath,amssymb,superscriptaddress,floatfix]{revtex4-2}

\usepackage[T1]{fontenc}
\usepackage[ansinew]{inputenc}
\usepackage{graphics}
\usepackage{dcolumn}
\usepackage{bm}
\usepackage{amsthm}
\usepackage{amsmath}
\usepackage{amssymb}
\usepackage{diagbox}
\usepackage{hyperref}
\usepackage{url}
\usepackage{xcolor}

\begin{document}
	
	\title{Magnetic microstructure of nanocrystalline Fe-Nb-B alloys as seen by small-angle neutron and X-ray scattering}
	
	\author{Venus Rai}
	\email{venus.rai@uni.lu}
	\affiliation{Department of Physics and Materials Science, University of Luxembourg, 162A avenue de la Faiencerie, L-1511 Luxembourg, Grand Duchy of Luxembourg}
	\author{Ivan Titov}
	\affiliation{Department of Physics and Materials Science, University of Luxembourg, 162A avenue de la Faiencerie, L-1511 Luxembourg, Grand Duchy of Luxembourg}
	\author{Michael P.\ Adams}
	\affiliation{Department of Physics and Materials Science, University of Luxembourg, 162A avenue de la Faiencerie, L-1511 Luxembourg, Grand Duchy of Luxembourg}
	\author{Kiyonori Suzuki}
	\affiliation{Department of Materials Science and Engineering, Monash University, Clayton, VIC 3800, Australia}
	\author{Joachim Kohlbrecher}
	\affiliation{Paul Scherrer Institute, CH-5232 Villigen PSI, Switzerland}
	\author{Andreas Michels}
	\email{andreas.michels@uni.lu}
	\affiliation{Department of Physics and Materials Science, University of Luxembourg, 162A avenue de la Faiencerie, L-1511 Luxembourg, Grand Duchy of Luxembourg}
	
	\begin{abstract}
		We have investigated the magnetic microstructure of two-phase Fe-Nb-B~based Nanoperm alloys using unpolarized small-angle neutron scattering (SANS) and small-angle X-ray scattering (SAXS). Our SANS analysis reveals a significantly large magnetic scattering contribution due to spin misalignment, primarily originating from the substantial jump in the longitudinal magnetization at the interfaces between the particles and the matrix. The magnetic scattering exhibits an angular anisotropy that resembles a clover-leaf-type pattern, consistent with the predictions of micromagnetic SANS theory. Analysis of the one-dimensional SANS data yields values for the exchange-stiffness constant and the average anisotropy and magnetostatic fields. The micromagnetic correlation lengths for all three samples exhibit a similar field variation with sizes ranging between about $10$$-$$35 \, \mathrm{nm}$. We also find that the nuclear and magnetic residual scattering component of the SANS cross section exhibits a similar $q$~dependency as the SAXS data. These findings further validate the applicability of the micromagnetic SANS theory, and the mesoscopic information obtained is important for the advancement of the soft magnetic properties of this class of material.
	\end{abstract}
	
	\maketitle
	
	\section{Introduction}
	
	Iron-based soft ferromagnetic alloys with an ultrafine-grained microstructure attract a lot of attention because of their high saturation magnetization, low coercivity, and low core losses (see, e.g., Refs.~\cite{yoshizawa1988new,herzer1990grain,herzer97,suzuki1999nanocrystalline,suzuki06,gutfleisch2011magne,makino2012,herzer13,hasiak2014impact,hasiak2019microstructure,ram2021soft,huang2023effect} and references therein). These properties result in their great potential for various technological applications, such as in turbine generators, high-frequency power transformers, interface transformers, or various consumer electronics devices. However, in order to build energy-efficient devices, core losses have to be minimized, which can be achieved by minimizing the magnetic anisotropy and increasing the saturation magnetization.
	
	Recent studies of soft magnetic Fe-B based Nano\-perm alloys have shown that, in addition to magnetic hysteresis and eddy current losses, significant core (power) losses in Nanoperm may arise due to magnetostriction as well~\cite{tsukahara2022role,kiyonori2024}. This excess loss is attributed to factors such as domain walls, inelastic lattice relaxation mediated by magnetostriction, and magnetic interactions, including spin misalignment due to longitudinal magnetization jumps. Understanding and enhancing functionality requires scrutinizing the grain and magnetic microstructures on a mesoscopic length scale, where many macroscopic magnetic properties are realized.
	
	Small-angle neutron scattering (SANS) emerges as a key experimental technique, offering insights into the magnetic microstructure within a range of $\sim$$1$$-$$1000 \, \mathrm{nm}$~\cite{muhlbauer2019magnetic,michels2021magnetic}. Due to the high penetrating power of neutrons, thick bulk samples can commonly be investigated. Moreover, the past two decades have witnessed a significant progress regarding the understanding of magnetic SANS, which enables the quantitative study of the mesocale magnetic microstructure of a wide range of nanocrystalline compounds (e.g., \cite{michels03epl,michels2006dipolar,suzuki2007,eckerlebe2011,bhatti2012,bergner2013,Pareja:15,grutter2017,shu2018,bender2018dipolar,mirebeau2018,mistonov2019,kons2020investigating,zakutna2020field,vivas2020toward,bersweiler2022unraveling,eli2023,arena2023,ukleev2024}).
	
	Here, we report SANS and small-angle X-ray scattering (SAXS) results on a series of Fe-based nanocrystalline alloys (Fe$_{87}$B$_{13}$, Fe$_{85}$Nb$_6$B$_{9}$, Fe$_{80}$Nb$_6$B$_{14}$). We observe a significant variation of the SANS cross section as the applied magnetic field is reduced from $9 \, \mathrm{T}$ down to about $0.1 \, \mathrm{T}$, which demonstrates the presence of a large spin-misalignment scattering in all the samples. The two-dimensional magnetic SANS signal exhibits an anisotropic clover-leaf-type pattern, in agreement with the prevalence of dipolar stray fields decorating the nanoparticles, in this way producing nanoscale spin disorder.
	
	The paper is organized as follows:~Section~\ref{exp} furnishes some details on the neutron experiment as well as on the structural and magnetic properties of the Nanoperm alloys. Section~\ref{sans_theory} sketches the basic ideas of the micromagnetic SANS theory, as it is implemented in the {\it MuMag2022} software tool~\cite{adams2022mumag2022} used for the neutron data analysis. Section~\ref{sans_sans_exp} features the neutron data and analysis, the ensuing results for the exchange constants and the average anisotropy and magnetostatic fields, and a comparison of the residual SANS cross section with laboratory SAXS data. Finally, in Sec.~\ref{summary} we summarize the main findings of this study. The Appendix displays the complete set of the two-dimensional total (nuclear and magnetic) and purely magnetic SANS cross sections of the Nanoperm alloys.
	
	\section{Experimental}
	\label{exp}
	
	\begin{table*}
		\small
		\centering
		\caption{Structural and magnetic properties of the investigated Nanoperm alloys. Data partially taken from Ref.~\cite{kiyonori2024}.}
		\begin{tabular}{lccccccc}
			\hline
			\hline
			Alloy & Annealing condition & Annealing temperature  & Thickness & Grain size  & $H_{\mathrm{c}}$~(A/m) & $\mu_0 M_0$~(T) & $\lambda_{\mathrm{s}}$~(ppm) 
			\tabularnewline[\doublerulesep]
			\hline
			\noalign{\vskip\doublerulesep}
			Fe$_{87}$B$_{13}$ & Ultra-rapid annealing & $753 \, \mathrm{K}$ (0.5~sec.) &  $13.8 \, \mu\mathrm{m}$ & $16 \, \mathrm{nm}$ &  7.3 & 1.89 & 13
			%	\tabularnewline[\doublerulesep]
			%	\hline
			\tabularnewline[\doublerulesep]
			Fe$_{85}$Nb$_6$B$_{9}$ & Tube annealing & $898 \, \mathrm{K}$ (30~min.) & $14.5 \, \mu\mathrm{m}$ & $11 \, \mathrm{nm}$ &  6.7 & 1.67 & $\sim$0
			\tabularnewline[\doublerulesep]
			Fe$_{80}$Nb$_6$B$_{14}$ & Tube annealing & $898 \, \mathrm{K}$ (30~min.) &  $24.5 \, \mu\mathrm{m}$ & $10 \, \mathrm{nm}$ &  3.9 & 1.50 & 2.4
			\tabularnewline[\doublerulesep]
			\hline
		\end{tabular}
		\label{tab:sample_char}
	\end{table*}
	
	Nanoperm samples with nominal compositions of Fe$_{87}$B$_{13}$, Fe$_{85}$Nb$_6$B$_{9}$, and Fe$_{80}$Nb$_6$B$_{14}$ were prepared by rapid solidification (melt spinning) followed by thermal annealing. This resulted in ultrafine-grained microstructures consisting of nanocrystalline Fe particles that are embedded in an amorphous Nb-B magnetic matrix. For the sample synthesis, the low neutron absorbing isotope $^{11}$B was used. The details of the sample preparation can be found in Refs.~\cite{li2020dramatic,kiyonori2024}. Further information on the annealing conditions during the synthesis procedure of each sample as well as the structural and magnetic properties are given in Table~\ref{tab:sample_char}.
	
\begin{figure}[tb!]
		\centering
		\resizebox{0.85\columnwidth}{!}{\includegraphics{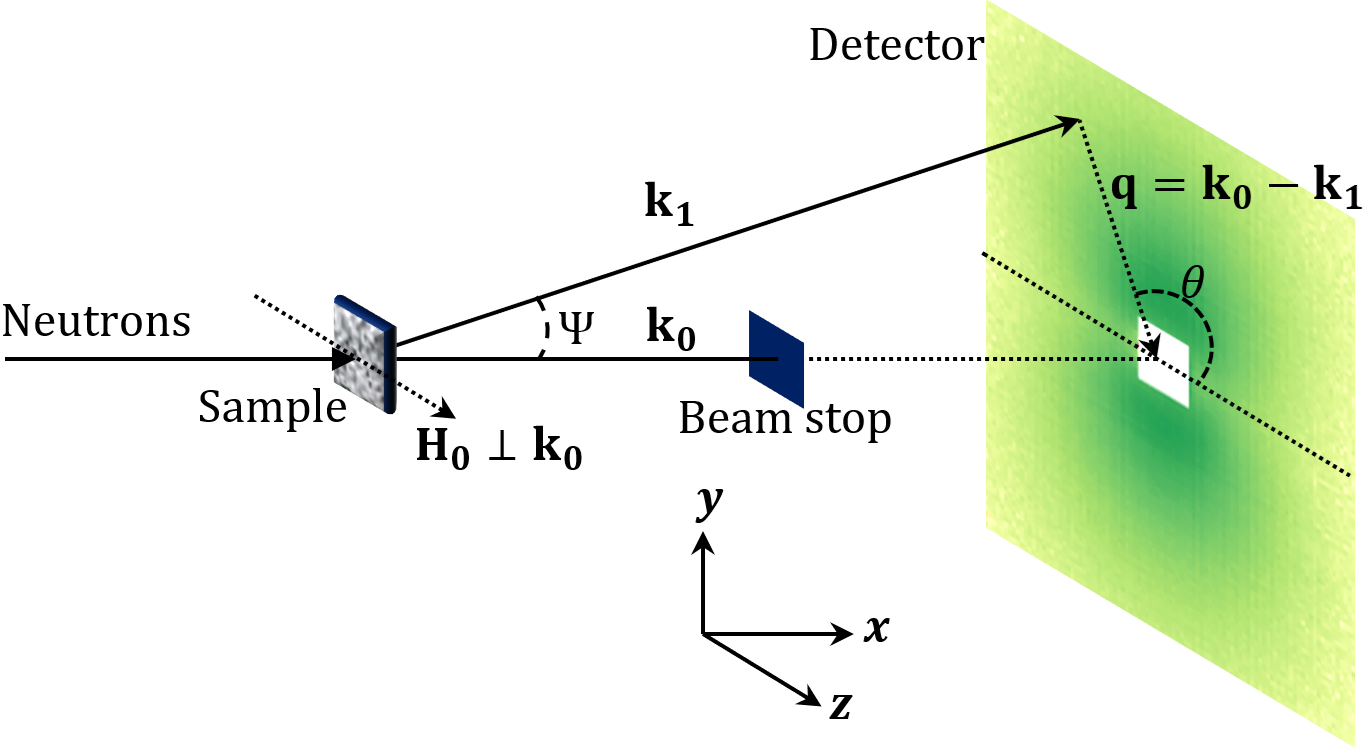}}
		\caption{Schematic drawing of the SANS setup. The wavevectors of the incoming and scattered neutron beams are, respectively, denoted by $\mathbf{k}_0$ and $\mathbf{k}_1$, and $\psi$ denotes the scattering angle. The momentum-transfer or scattering vector equals $\mathbf{q} = \mathbf{k}_0 - \mathbf{k}_1$. The azimuthal angle $\theta$ describes the angular anisotropy of the scattered neutron intensity on the 2D detector. The direction of the applied magnetic field $\mathbf{H}_0$ is horizontal and perpendicular to the incident beam. In our notation, $\mathbf{H}_0$ defines the $z$~axis of a Cartesian global frame, the incoming beam ($\mathbf{k}_0$) is along the $x$~axis, and the vertical direction defines the $y$~axis.}
		\label{fig:neutron_scatt}
	\end{figure}

	The unpolarized SANS experiment was performed using the SANS-I instrument at the Swiss Spallation Neutron Source at the Paul Scherrer Institute, Switzerland. A schematic drawing of the SANS experiment is depicted in Fig.~\ref{fig:neutron_scatt}. We used an incident mean neutron wavelength of $\lambda = 6$~{\AA} with a wavelength broadening of $\Delta\lambda/\lambda = 10\,\%$ (FWHM). For the SANS experiment, typically $10$$-$$15$ sheets of Nanoperm were stacked together, resulting in total sample thicknesses of $\sim$$150$$-$$250 \, \mu\mathrm{m}$ (compare to Table~\ref{tab:sample_char}). The entire SANS experiment was conducted at room temperature and under the application of an external magnetic field ($\mu_0 H_0^{\mathrm{max}} = 9 \, \mathrm{T}$), applied along the horizontal direction and perpendicular to the incoming neutron beam (compare to Fig.~\ref{fig:neutron_scatt}). The SANS intensity for each sample was recorded at three different sample-to-detector distances ($2 \, \mathrm{m}$, $6 \, \mathrm{m}$, $18 \, \mathrm{m}$), enabling data collection in a $q$~range of $\sim$$0.03$$-$$3 \, \mathrm{nm}^{-1}$. The neutron data reduction (corrections for background scattering and sample transmission) was conducted using the GRASP software package~\cite{dewhurst2023}.
	
	Wide-angle X-ray diffraction measurements of all the samples were performed in a Bruker D8 DISCOVER diffractometer setup in Bragg-Brentano geometry using Cu-K$_{\alpha}$ radiation ($\lambda = 0.154 \, \mathrm{nm}$). Characteristic Bragg peaks corresponding to the pure Fe phase (BCC) were observed. The full-width at half maxima (FWHM) of these peaks was analyzed and the average crystallite size was estimated using the Scherrer equation. The obtained particle sizes (grain sizes) are given in Table~\ref{tab:sample_char}. Furthermore, all three samples were characterized by small-angle X-ray scattering (SAXS) using an Anton Paar SAXSpoint 5.0 instrument. Monochromatic X-rays with a wavelength of $\lambda = 0.154 \, \mathrm{nm}$ were used to investigate the Nanoperm samples at scattering vectors $q = 4\pi \sin(\psi/2) / \lambda$ ranging from about $0.045 \, \mathrm{nm}^{-1}$ to $2.1 \, \mathrm{nm}^{-1}$, where $\psi$ denotes the scattering angle (compare to Fig.~\ref{fig:neutron_scatt}).
	
\begin{figure}[tb!]
		\centering
		\resizebox{0.90\columnwidth}{!}{\includegraphics{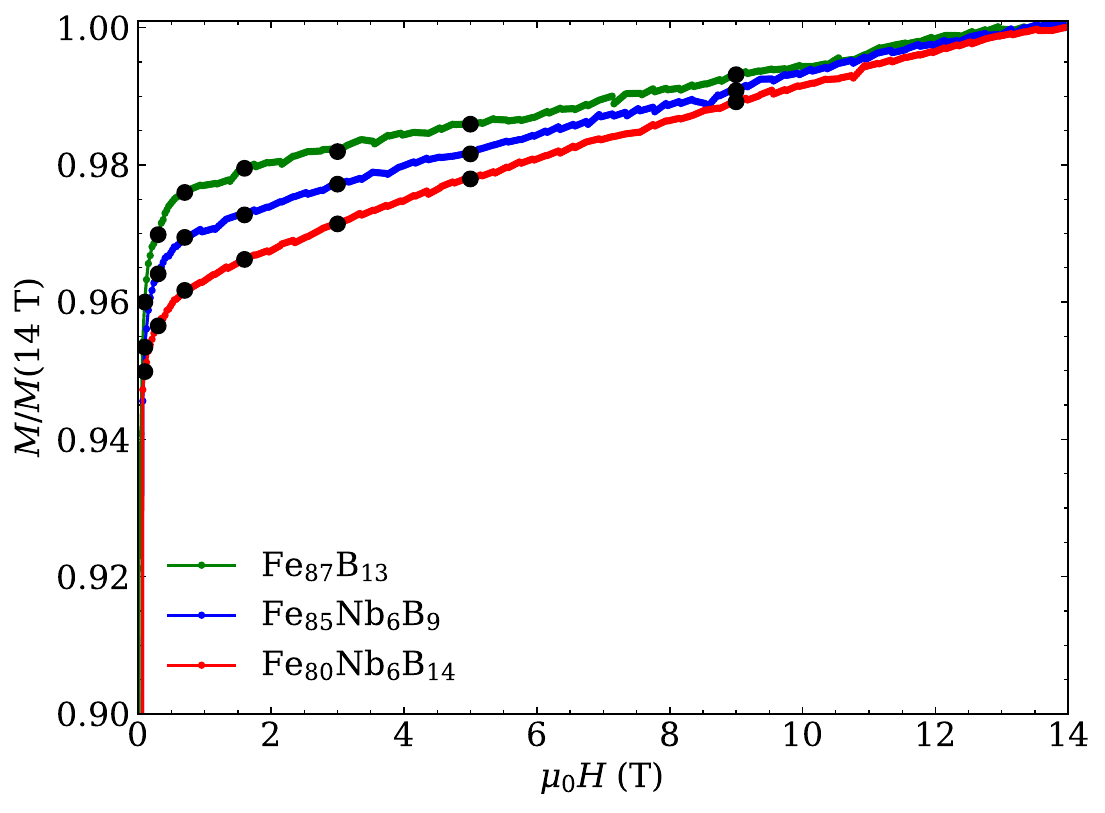}}
		\caption{Normalized room-temperature magnetization curves of the Nanoperm samples (see inset; only the upper right quadrant is shown). The black dots mark the field values of the SANS measurements.}
		\label{fig:mh}
	\end{figure}

	Room temperature magnetization measurements were performed using a Cryogenics Ltd.\ vibrating sample magnetometer setup, equipped with a $14 \, \mathrm{T}$ superconducting magnet. The normalized magnetization data are displayed in Fig.~\ref{fig:mh}. As mentioned in Table~\ref{tab:sample_char}, the saturation magnetizations of all three samples are significantly different~\cite{kiyonori2024}. However, the approach-to-saturation regime [$M/M(14 \, \mathrm{T}) \gtrsim 95 \,\%$] is, for all three samples, attained for applied fields larger than about $0.1 \, \mathrm{T}$. This should guarantee the validity of the micromagnetic SANS analysis. Magnetic parameters are listed in Table~\ref{tab:sample_char}.
	
	\section{Magnetic SANS Theory}
	\label{sans_theory}
	
	The total unpolarized SANS cross section of a bulk ferromagnet consists of nuclear and magnetic scattering contributions. The origin of magnetic SANS are mesoscale spatial variations in the magnitude and orientation of the magnetization vector field $\mathbf{M}(\mathbf{r}) = \{ M_x(\mathbf{r}), M_y(\mathbf{r}), M_z(\mathbf{r}) \}$, where $M_z$ denotes the longitudinal magnetization (parallel to $\mathbf{H}_0$) and $M_{x,y}$ are the two transversal components~\cite{muhlbauer2019magnetic,michels2021magnetic}. The Nanoperm alloys studied in this paper are magnetically extremely soft, implying that a very small field of the order of a few mT will bring the material close to magnetic saturation. In this case the magnetic small-angle scattering due to the $M_z$~fluctuations can be approximated to be independent of the applied magnetic field, while the magnetic SANS due to the $M_{x,y}$, called spin-misalignment scattering, is strongly field-dependent, in particular at small momentum transfers. The micromagnetic SANS theory that is here used to analyze the Nanoperm alloys is detailed in Refs.~\cite{honecker2013theory,honecker2013analysis} and implemented in the {\it MuMag2022} software tool~\cite{adams2022mumag2022}. In the following, we recall the basic expressions to achieve a self-contained presentation.
	
	As shown in Ref.~\cite{honecker2013theory}, near magnetic saturation, the total unpolarized SANS cross section $d \Sigma / d \Omega$ can be evaluated by means of micromagnetic theory. In particular,
	\begin{equation}
		\label{sigmasansperp2d}
		\frac{d \Sigma}{d \Omega}(\mathbf{q}) = \frac{d \Sigma_{\mathrm{res}}}{d \Omega}(\mathbf{q}) + \frac{d \Sigma_{\mathrm{M}}}{d \Omega}(\mathbf{q}) ,
	\end{equation}
	where
	\begin{equation}
		\label{sigmaresperp1}
		\frac{d \Sigma_{\mathrm{res}}}{d \Omega}(\mathbf{q}) = \frac{8 \pi^3}{V} \left( |\widetilde{N}|^2
		+ b_{\mathrm{H}}^2 |\widetilde{M}_z|^2 \sin^2\theta \right)
	\end{equation}
	represents the nuclear and magnetic residual SANS cross section, which is measured at complete magnetic saturation (infinite field), and
	\begin{equation}
		\label{sigmasmperp}
		\frac{d \Sigma_{\mathrm{M}}}{d \Omega}(\mathbf{q}) = S_{\mathrm{H}}(\mathbf{q}) \, R_{\mathrm{H}}(q, \theta, H_{\mathrm{i}}) + S_{\mathrm{M}}(\mathbf{q}) \, R_{\mathrm{M}}(q, \theta, H_{\mathrm{i}})
	\end{equation}
	is the spin-misalignment SANS cross section. In Eq.~(\ref{sigmaresperp1}), $V$ is the scattering volume, $b_{\mathrm{H}} = 2.91 \times 10^{8} \, \mathrm{A^{-1} m^{-1}}$ is the magnetic scattering length, $\widetilde{N}(\mathbf{q})$ and $\widetilde{\mathbf{M}}(\mathbf{q}) = \{ \widetilde{M}_x(\mathbf{q}), \widetilde{M}_y(\mathbf{q}), \widetilde{M}_z(\mathbf{q}) \}$ denote, respectively, the Fourier transforms of the nuclear scattering-length density and of the magnetization $\mathbf{M}(\mathbf{r})$, and $\theta$ represents the angle between $\mathbf{H}_0$ and $\mathbf{q}$ (see Fig.~\ref{fig:neutron_scatt}). The magnetic scattering due to transversal spin components, with related Fourier amplitudes $\widetilde{M}_x(\mathbf{q})$ and $\widetilde{M}_y(\mathbf{q})$, is contained in $d \Sigma_{\mathrm{M}} / d \Omega$, which decomposes into a contribution $S_{\mathrm{H}} R_{\mathrm{H}}$ due to perturbing magnetic anisotropy fields and a part $S_{\mathrm{M}} R_{\mathrm{M}}$ related to magnetostatic fields. The micromagnetic SANS theory considers a uniform exchange interaction and a random distribution of the magnetic easy axes, as it is appropriate for a statistically-isotropic polycrystalline ferromagnet~\cite{michels2021magnetic}. Spatial variations in the magnitude of the saturation magnetization are explicitly taken into account via the function $S_{\mathrm{M}}$ (see below). Moreover, in the approach-to-saturation regime it is assumed that $|\widetilde{M}_z|^2 = |\widetilde{M}_{\mathrm{s}}|^2$, where $\widetilde{M}_{\mathrm{s}}(\mathbf{q})$ denotes the Fourier transform of the saturation magnetization profile $M_{\mathrm{s}}(\mathbf{r})$.
	
	Regarding the decomposition of the SANS cross section [Eq.~(\ref{sigmasansperp2d})], it is important to emphasize that it is $d \Sigma_{\mathrm{M}} / d \Omega$ which depends on the magnetic interactions (exchange, anisotropy, magnetostatics), while $d \Sigma_{\mathrm{res}} / d \Omega$ is determined by the geometry of the underlying grain microstructure (e.g., the particle shape or the particle-size distribution). If in a SANS experiment the approach-to-saturation regime can be reached for a particular magnetic material (as it is the case for the Nanoperm alloys), then the residual SANS can be obtained by an analysis of field-dependent data via the extrapolation to infinite field. In a sense, for a bulk ferromagnet, the scattering at saturation resembles the topographical background in Kerr-microscopy experiments, which needs to be subtracted in order to access the magnetic domain structure of the sample~\cite{mccord1999}.
	
	The anisotropy-field scattering function (in units of $\mathrm{cm}^{-1}$)
	\begin{equation}
		\label{shdef}
		S_{\mathrm{H}}(\mathbf{q}) = \frac{8 \pi^3}{V} \, b_{\mathrm{H}}^2 \, |\mathbf{\widetilde{H}}_{\mathrm{p}}|^2
	\end{equation}
	depends on $\mathbf{\widetilde{H}}_{\mathrm{p}}(\mathbf{q})$, which represents the Fourier transform of the spatial structure of the magnetic anisotropy field $\mathbf{H}_{\mathrm{p}}(\mathbf{r})$ of the sample, whereas the scattering function of the longitudinal magnetization (in units of $\mathrm{cm}^{-1}$)
	\begin{equation}
		\label{smdef}
		S_{\mathrm{M}}(\mathbf{q}) = \frac{8 \pi^3}{V} \, b_{\mathrm{H}}^2 \, |\widetilde{M}_z|^2
	\end{equation}
	provides information on the spatial variation of the saturation magnetization $M_{\mathrm{s}}(\mathbf{r})$; for instance, in a multiphase magnetic nanocomposite, $S_{\mathrm{M}} \propto |\widetilde{M}_z|^2 \propto (\Delta M)^2$, where $\Delta M$ denotes the jump of the magnetization magnitude at internal (particle-matrix) interfaces. Note that the volume average of $M_{\mathrm{s}}(\mathbf{r})$ equals the macroscopic saturation magnetization $M_0 = \langle M_{\mathrm{s}}(\mathbf{r}) \rangle$ of the sample, which can be measured with a magnetometer. The corresponding dimensionless micromagnetic response functions can be expressed as~\cite{honecker2013theory}:
	\begin{equation}
		\label{rhdefperp}
		R_{\mathrm{H}}(q, \theta, H_{\mathrm{i}}) = \frac{p^2}{2} \left( 1 + \frac{\cos^2\theta}{\left( 1 + p \sin^2\theta \right)^2} \right)
	\end{equation}
	and
	\begin{equation}
		\label{rmdefperp}
		R_{\mathrm{M}}(q, \theta, H_{\mathrm{i}}) = \frac{p^2 \, \sin^2\theta \cos^4\theta}{\left( 1 + p \sin^2\theta \right)^2} + \frac{2 p \, \sin^2\theta \cos^2\theta}{1 + p \sin^2\theta} ,
	\end{equation}
	where
	\begin{equation}
		\label{pdef}
		p(q, H_{\rm i}) = \frac{M_0}{H_{\rm eff}(q, H_{\mathrm{i}})} 
	\end{equation}
	is a dimensionless function, and $\theta$ represents the angle between $\mathbf{H}_0 = H_0 \mathbf{e}_z$ and $\mathbf{q} \cong q \{ 0, \sin\theta, \cos\theta \}$. The effective magnetic field
	\begin{equation}
		\label{heffdef}
		H_{\mathrm{eff}}(q, H_{\mathrm{i}}) = H_{\mathrm{i}} \left( 1 + l_{\mathrm{H}}^2 q^2 \right) = H_{\mathrm{i}} + \frac{2A}{\mu_0 M_0} q^2
	\end{equation}
	depends on the internal magnetic field
	\begin{equation}
		\label{hidef}
		H_{\mathrm{i}} = H_0 - H_{\mathrm{d}} = H_0 - N_{\mathrm{d}} M_0 > 0
	\end{equation}
	and on the micromagnetic exchange length of the field
	\begin{equation}
		\label{lhdef}
		l_{\mathrm{H}}(H_{\mathrm{i}}) = \sqrt{\frac{2 A}{\mu_0 M_0 H_{\mathrm{i}}}}
	\end{equation}
	($M_0$:~saturation magnetization; $A$:~exchange-stiffness parameter; $H_{\mathrm{d}} = N_{\mathrm{d}} M_0$:~demagnetizing field; $0 \leq N_{\mathrm{d}} \leq 1$:~demagnetizing factor; $\mu_0 = 4\pi 10^{-7} \, \mathrm{T m/A}$). Note that $H_0 \gg H_{\mathrm{d}}$ in the approach-to-saturation regime. The $\theta$-dependence of $R_{\mathrm{H}}$ and $R_{\mathrm{M}}$ arises essentially as a consequence of the magnetodipolar interaction. Depending on the values of $q$ and $H_{\mathrm{i}}$, a variety of angular anisotropies may be seen on a two-dimensional position-sensitive detector~\cite{michels2021magnetic}.
	
	The effective magnetic field $H_{\mathrm{eff}}$ [Eq.~(\ref{heffdef})] consists of a contribution due to the internal field $H_{\mathrm{i}}$ and of the exchange field $2 A q^2/ (\mu_0 M_{\mathrm{0}})$. An increase of $H_{\mathrm{i}}$ increases the effective field only at the smallest $q$~values, whereas $H_{\mathrm{eff}}$ at the larger $q$ is always very large ($\sim$$10$$-$$100 \, \mathrm{T}$) and independent of $H_{\mathrm{i}}$~\cite{michels2021magnetic}. The latter statement may be seen as a manifestation of the fact that exchange forces tend to dominate on small length scales~\cite{aharonibook}. The role of $H_{\mathrm{eff}}$ is to suppress the high-$q$ Fourier components of the magnetization, which correspond to sharp real-space fluctuations. On the other hand, long-range magnetization fluctuations, at small $q$, are effectively suppressed when $H_{\mathrm{i}}$ is increased.
	
	By assuming the functions $\widetilde{N}$, $\widetilde{M}_z$, and $\mathbf{\widetilde{H}}_{\mathrm{p}}$ to depend only on the magnitude $q = | \mathbf{q} |$ of the scattering vector, one can perform an azimuthal average of Eq.~(\ref{sigmasansperp2d}), i.e., $1/(2\pi) \int_0^{2\pi} (...) d\theta$. The resulting expressions for the response functions then read:
	\begin{equation}
		\label{rhdefperpradav}
		R_{\mathrm{H}}(q, H_{\mathrm{i}}) = \frac{p^2}{4} \left( 2 + \frac{1}{\sqrt{1 + p}} \right)
	\end{equation}
	and
	\begin{equation}
		\label{rmdefperpradav}
		R_{\mathrm{M}}(q, H_{\mathrm{i}}) = \frac{\sqrt{1 + p} - 1}{2} ,
	\end{equation}
	so that the azimuthally-averaged total nuclear and magnetic SANS cross section can be written as:
	\begin{eqnarray}
		\label{sigmafinalperp}
		\frac{d \Sigma}{d \Omega}(q, H_{\mathrm{i}}) &=& \frac{d \Sigma_{\mathrm{res}}}{d \Omega}(q) + \frac{d \Sigma_{\mathrm{M}}}{d \Omega}(q, H_{\mathrm{i}}) \\
		&=& \frac{d \Sigma_{\mathrm{res}}}{d \Omega}(q) + S_{\mathrm{H}}(q) \, R_{\mathrm{H}}(q, H_{\mathrm{i}}) \nonumber \\
		& & + S_{\mathrm{M}}(q) \, R_{\mathrm{M}}(q, H_{\mathrm{i}}) \nonumber ,
	\end{eqnarray}
	where
	\begin{equation}
		\label{sigmaresperp}
		\frac{d \Sigma_{\mathrm{res}}}{d \Omega}(q) = \frac{8 \pi^3}{V} \left( |\widetilde{N}(q)|^2
		+ \frac{1}{2} \, b_{\mathrm{H}}^2 \, |\widetilde{M}_z(q)|^2 \right) .
	\end{equation}
	For materials exhibiting a uniform saturation magnetization (e.g., single-phase materials), the magnetostatic scattering contribution $S_{\mathrm{M}} R_{\mathrm{M}}$ to $d \Sigma_{\mathrm{M}} / d \Omega$ [compare Eq.~(\ref{sigmasmperp})] is expected to be much smaller than the anisotropy-field related term $S_{\mathrm{H}} R_{\mathrm{H}}$~(compare, e.g., Fig.~23 in Ref.~\cite{michels2014review}).
	
	We emphasize that the micromagnetic theory behind the {\it MuMag2022} software results in an analytical expression for the two-dimensional SANS cross section as a function of the magnitude $q$ and the orientation $\theta$ of the scattering vector $\mathbf{q}$. These analytical expressions can be azimuthally-averaged over the full angular detector range $2\pi$ (or any other range) and compared to correspondingly-averaged experimental SANS data, in other words, it is not required that the experimental input SANS data are isotropic. Equation~(\ref{sigmafinalperp}) is the central expression that is used here to analyze the 1D SANS data. The free parameters are $d \Sigma_{\mathrm{res}} / d \Omega$, $S_{\mathrm{H}}$, $S_{\mathrm{M}}$, and the exchange constant $A$ in the expressions for $R_{\mathrm{H}}$ and $R_{\mathrm{M}}$. Numerical integration of the obtained $S_{\mathrm{H}}(q)$ and $S_{\mathrm{M}}(q)$ over the whole $\mathbf{q}$-space, i.e.,
	\begin{eqnarray}
		\label{Hpvoldef}
		\langle |\mathbf{H}_{\mathrm{p}}|^2 \rangle = \frac{1}{2\pi^2 b_{\mathrm{H}}^2} \int_0^{\infty} S_{\mathrm{H}}(q) \, q^2 \, dq , \\
		\label{Mzvoldef}
		\langle |M_z|^2 \rangle = \frac{1}{2\pi^2 b_{\mathrm{H}}^2} \int_0^{\infty} S_{\mathrm{M}}(q) \, q^2 \, dq ,
	\end{eqnarray}
	yields, respectively, the mean-square anisotropy field $\langle |\mathbf{H}_{\mathrm{p}}|^2 \rangle$ and the mean-square longitudinal magnetization fluctuation $\langle |M_z|^2 \rangle$~\cite{michels2021magnetic}. Since experimental data for $S_{\mathrm{H}}$ and $S_{\mathrm{M}}$ are only available within a finite range of momentum transfers between $q_{\mathrm{min}}$ and $q_{\mathrm{max}}$, one can only obtain rough lower bounds for these quantities. Therefore, the numerical integrations in Eqs.~(\ref{Hpvoldef}) and (\ref{Mzvoldef}) are carried out for $q_{\mathrm{min}} \leq q \leq q_{\mathrm{max}}$; $q_{\mathrm{min}}$ denotes the first experimental data point, while $q_{\mathrm{max}} \cong [\mu_0 M_0 H_{\mathrm{max}} /(2A)]^{1/2}$ can be estimated based on the value of the maximum applied magnetic field $H_{\mathrm{max}}$. For $q \gtrsim q_{\mathrm{max}}$, the reliable separation of the spin-misalignment ($S_{\mathrm{H}} R_{\mathrm{H}} + S_{\mathrm{M}} R_{\mathrm{M}}$) and residual scattering ($d \Sigma_{\mathrm{res}} / d \Omega$) is difficult (since then one attempts to fit a straight line to a constant), and the micromagnetic analysis should therefore be restricted to $q \lesssim q_{\mathrm{max}}$; see Ref.~\cite{adams2022mumag2022} for further details on the data analysis and the fit procedure.
	
	\begin{figure*}
		\centering
		\resizebox{1.65\columnwidth}{!}{\includegraphics{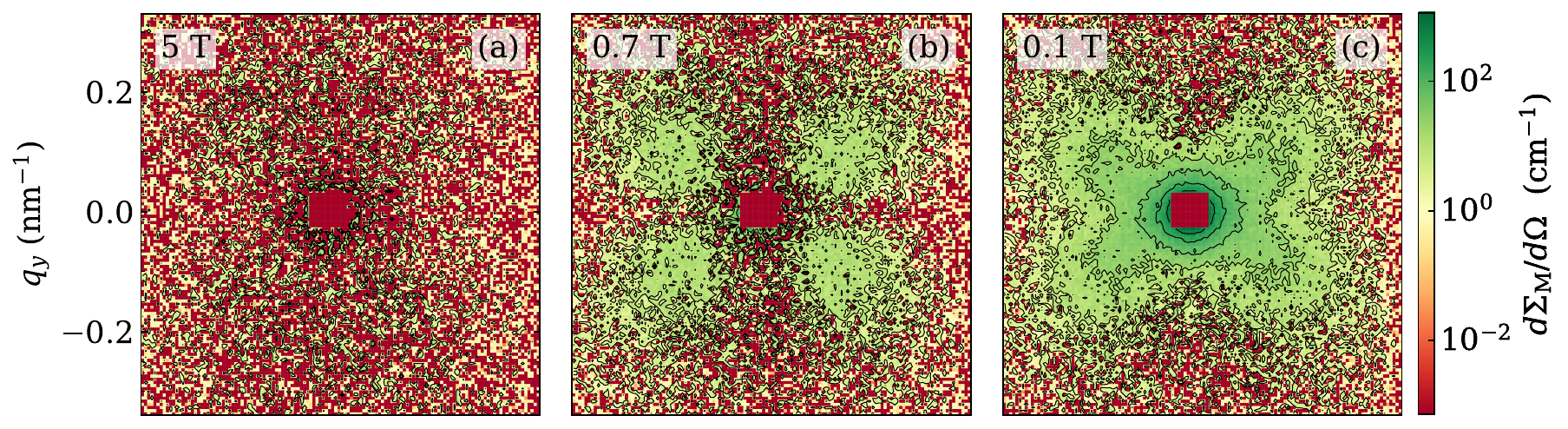}} \\
		\resizebox{1.65\columnwidth}{!}{\includegraphics{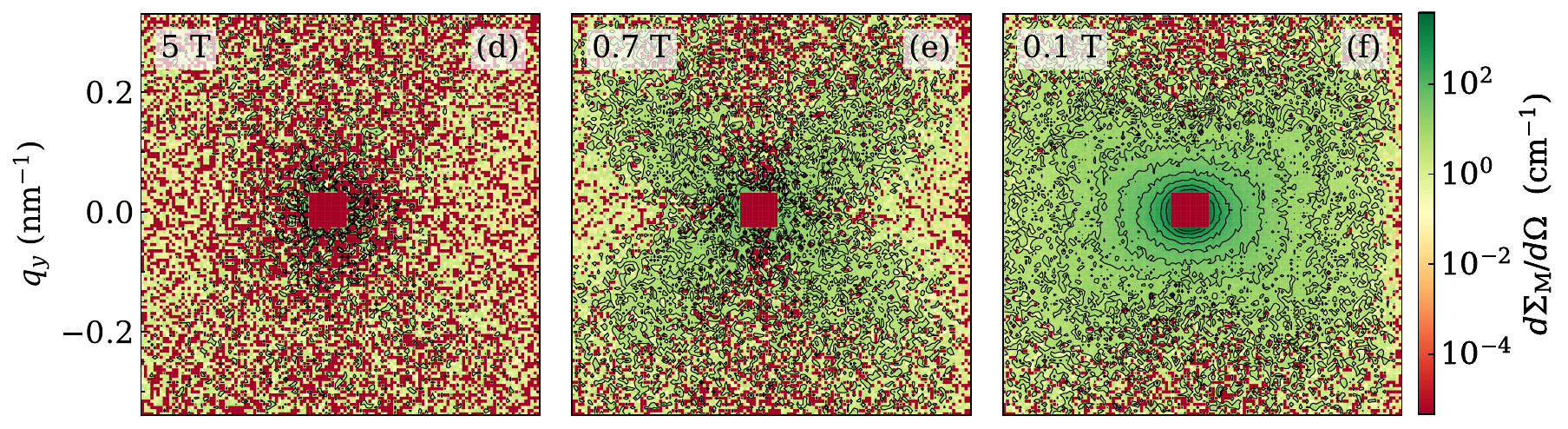}} \\
		\resizebox{1.65\columnwidth}{!}{\includegraphics{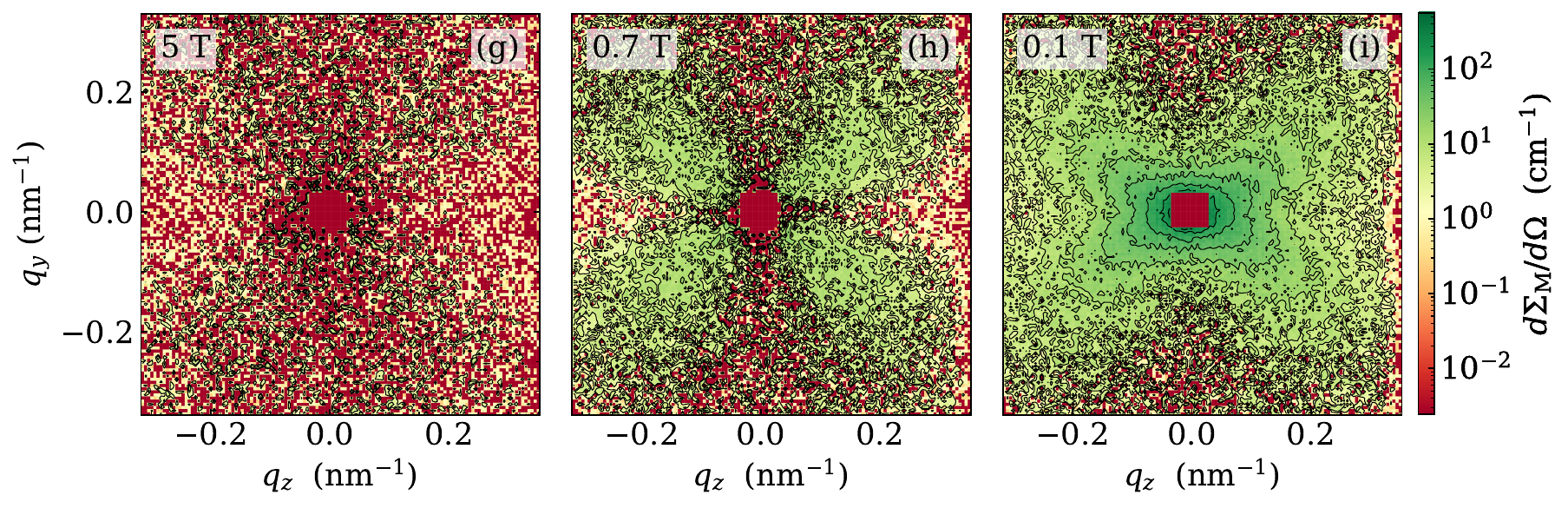}}
		\caption{Magnetic SANS cross sections of Nanoperm alloys (sample-to-detector distance:\ $18 \, \mathrm{m}$) (logarithmic color scale). The data were obtained by subtracting the respective total (nuclear and magnetic) SANS intensity at $9 \, \mathrm{T}$ from the measurements at the lower fields (indicated in the top left corner of each subfigure). $\mathbf{H}_0 \parallel \mathbf{e}_z$ is horizontal. (a)$-$(c) Fe$_{87}$B$_{13}$; (d)$-$(f) Fe$_{85}$Nb$_6$B$_{9}$; (g)$-$(i) Fe$_{80}$Nb$_6$B$_{14}$.}
		\label{fig:SANS_mag_selected}
	\end{figure*}
	
	\begin{figure*}[h]
		\centering
		\resizebox{1.70\columnwidth}{!}{\includegraphics{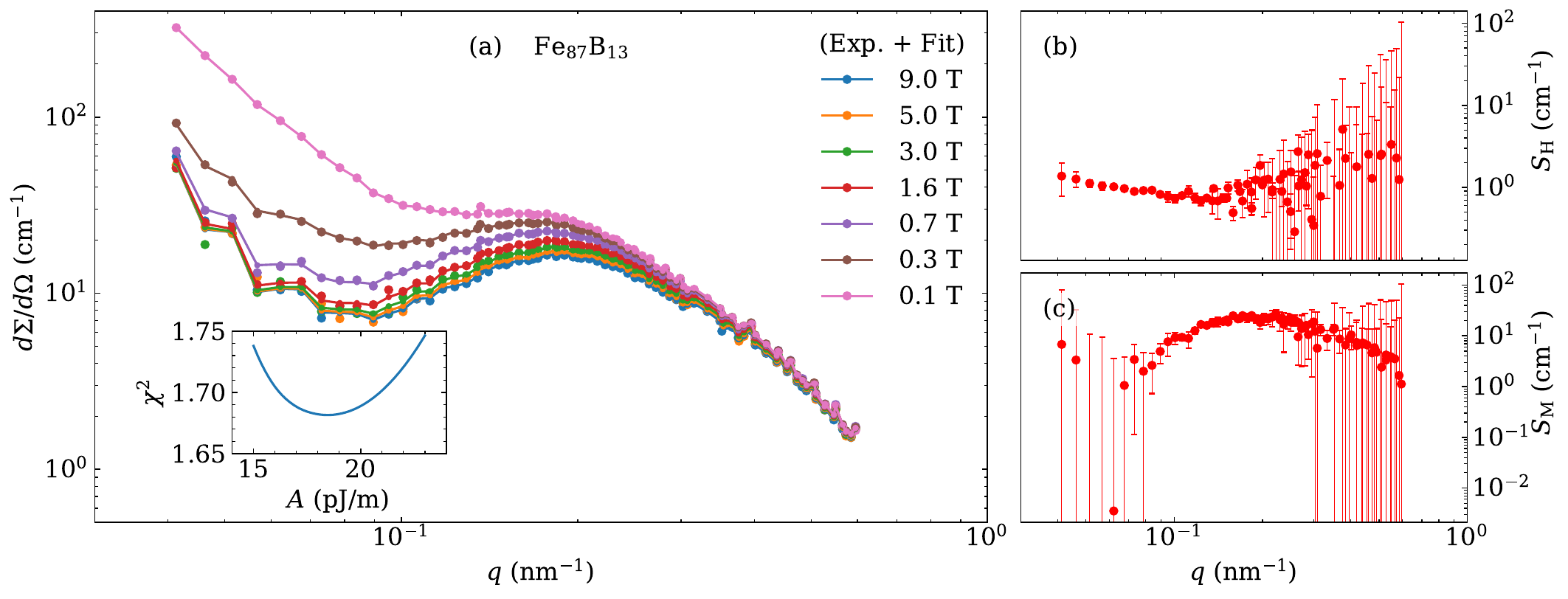}} \\
		\resizebox{1.70\columnwidth}{!}{\includegraphics{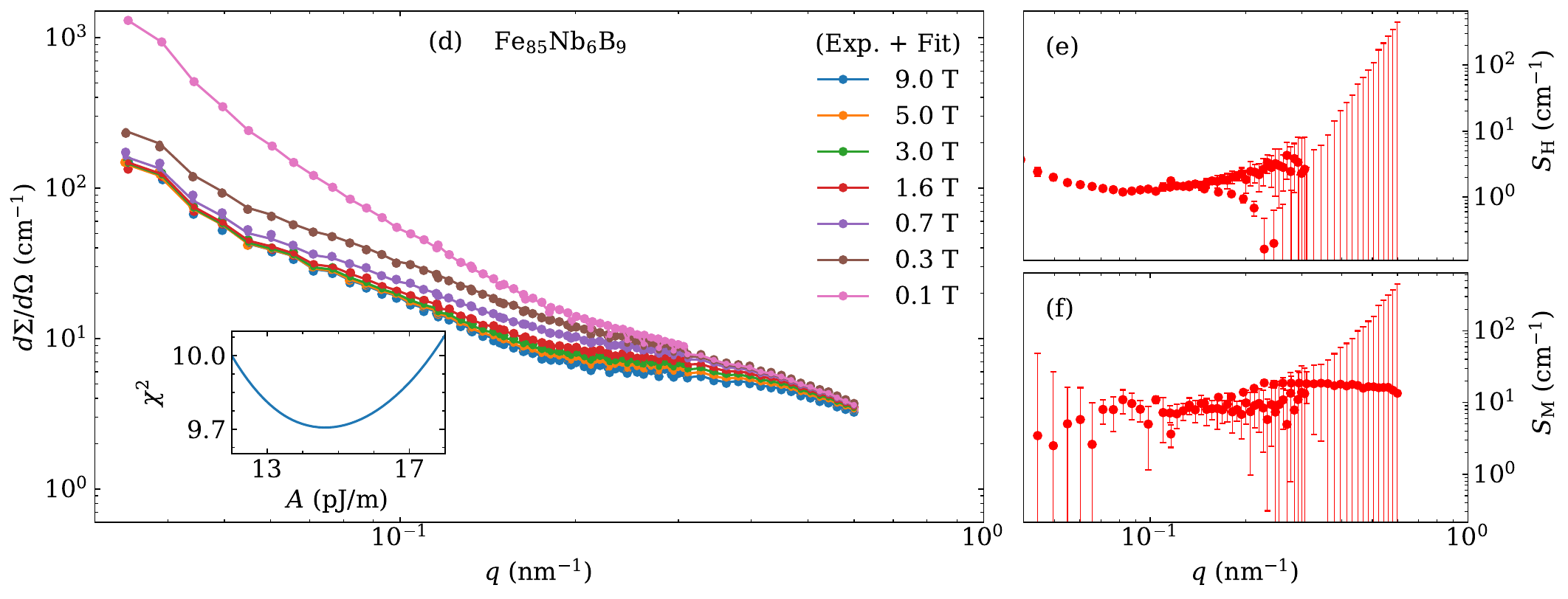}} \\
		\resizebox{1.70\columnwidth}{!}{\includegraphics{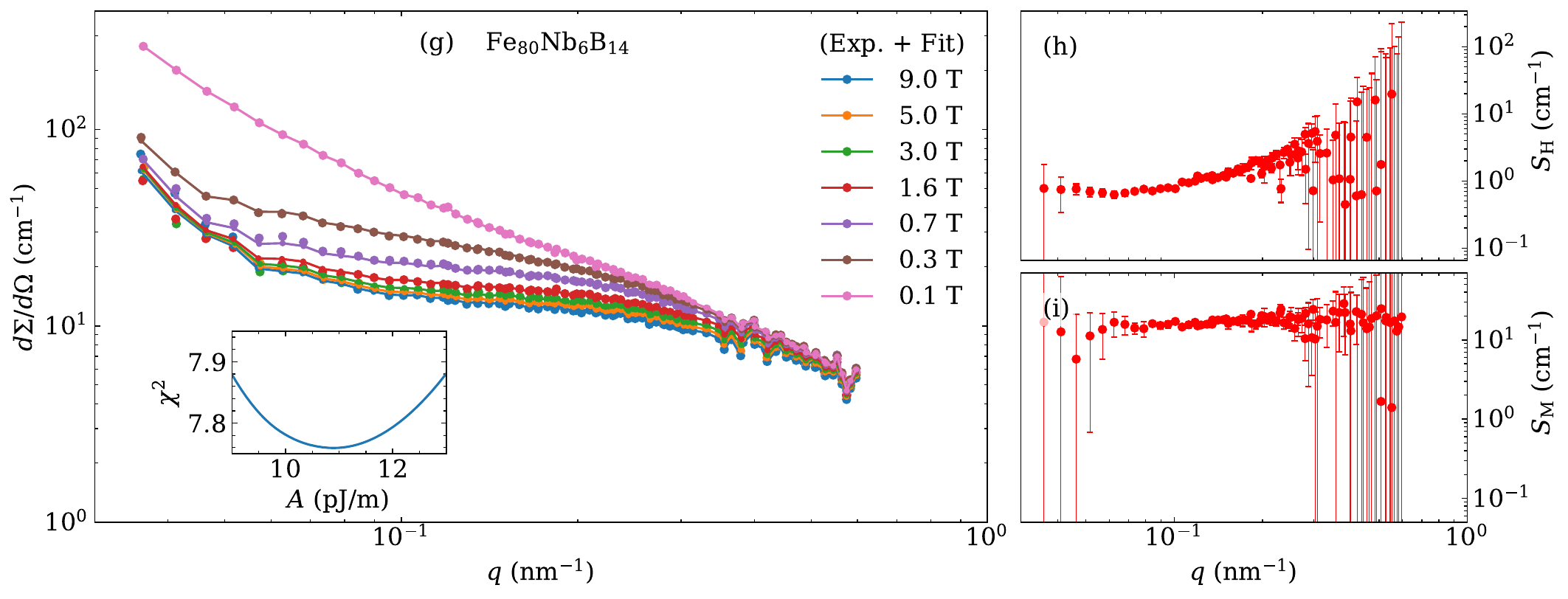}}
		\caption{Results of the micromagnetic SANS data analysis for the azimuthally-averaged total unpolarized SANS cross section. (a)$-$(c) Fe$_{87}$B$_{13}$; (d)$-$(f) Fe$_{85}$Nb$_6$B$_{9}$; (g)$-$(i) Fe$_{80}$Nb$_6$B$_{14}$. In (a, d, g), the dots denote the respective experimental data and the solid lines connect the computed cross-section values at the discrete $q$ and $H_{\mathrm{i}}$ using Eq.~(\ref{sigmafinalperp}). Insets in (a, d, g) show the goodness of fit, $\chi^2$, for a range of exchange-stiffness constants $A$. Subfigures (b,c), (e,f), and (h,i) display the respective results for the anisotropy-field ($S_{\mathrm{H}}$) and magnetostatic ($S_{\mathrm{M}}$) scattering functions. The fit analysis was restricted to $q \lesssim q_{\mathrm{max}} = 0.6 \, \mathrm{nm}^{-1}$ for all samples.}
		\label{fig:sans_1d_analysis}
	\end{figure*}
	
	\section{Results and Discussion}
	\label{sans_sans_exp}
	
	Figure~\ref{fig:SANS_mag_selected} displays, for the three alloy systems studied, the two-dimensional purely magnetic SANS cross section $d \Sigma_{\mathrm{M}} / d \Omega$ at selected applied magnetic fields, obtained after subtracting the respective nuclear and magnetic scattering at a saturating field of $9 \, \mathrm{T}$. The Appendix features a summary of the neutron data for the total $d \Sigma / d \Omega$ and $d \Sigma_{\mathrm{M}} / d \Omega$ at a number of fields and for all three sample-to-detector distances. As discussed in the previous section, the subtraction procedure approximately removes the nuclear and (longitudinal) magnetic scattering due to a saturated microstructure and in this way highlights the magnetic scattering that is related to the magnetic interactions. With decreasing field we note in Fig.~\ref{fig:SANS_mag_selected} (and in the corresponding $d \Sigma_{\mathrm{M}} / d \Omega$ data sets in the Appendix) the appearance of a pronounced angular anisotropy in $d \Sigma_{\mathrm{M}} / d \Omega$ with maxima roughly along the diagonals of the detector. Comparison to Eqs.~(\ref{sigmasmperp}) and (\ref{rmdefperp}) suggests that this so-called clover-leaf anisotropy is related to the magnetostatic scattering term $S_{\mathrm{M}} R_{\mathrm{M}}$ rather than to the anisotropy-field related contribution $S_{\mathrm{H}} R_{\mathrm{H}}$. As we will see below, the clover leaf pattern is very likely related to nanoscale jumps in the magnetization magnitude at internal particle-matrix interfaces, which give rise to local magnetostatic stray fields causing spin disorder. Similar angular anisotropies have also been reported for other compounds~\cite{michels2006dipolar,ogrin2006micromagnetic,michels2005dipole,michels2007temperature,michels08epl,elmas09,bersweiler2022unraveling,muhlbauer2019magnetic}.
	
	Figure~\ref{fig:sans_1d_analysis} shows the results of the micromagnetic SANS data analysis for the azimuthally-averaged {\it total} unpolarized SANS cross section $d \Sigma / d \Omega$. The analysis, carried out using the {\it MuMag2022} software~\cite{adams2022mumag2022}, has been restricted to applied fields larger than about $0.1 \, \mathrm{T}$, where the normalized magnetization is larger than $\sim$$95 \, \%$, so that all the Nanoperm samples are in the approach-to-saturation regime (compare Fig.~\ref{fig:mh}). It is seen that the computed cross sections based on the micromagnetic SANS theory [solid lines, Eq.~(\ref{sigmafinalperp})] very well reproduce the experimental data. As is typical for spin-misalignment scattering of bulk ferromagnets, the largest field dependence of $d \Sigma / d \Omega$ appears at the smallest momentum transfers, where also the clover-leaf pattern in the 2D data is most pronounced. The SANS data analysis has been restricted to $q \lesssim q_{\mathrm{max}} = 0.6 \, \mathrm{nm}^{-1}$, since the reliable separation of the spin-misalignment ($S_{\mathrm{H}} R_{\mathrm{H}} + S_{\mathrm{M}} R_{\mathrm{M}}$) and residual scattering ($d \Sigma_{\mathrm{res}} / d \Omega$) is difficult for $q \gtrsim q_{\mathrm{max}}$. This explains the increase of the uncertainty values in the data for $S_{\mathrm{H}}$ and $S_{\mathrm{M}}$ in Fig.~\ref{fig:sans_1d_analysis} with increasing momentum transfer $q$.
	
\begin{figure}[h]
		\centering
		\resizebox{0.90\columnwidth}{!}{\includegraphics{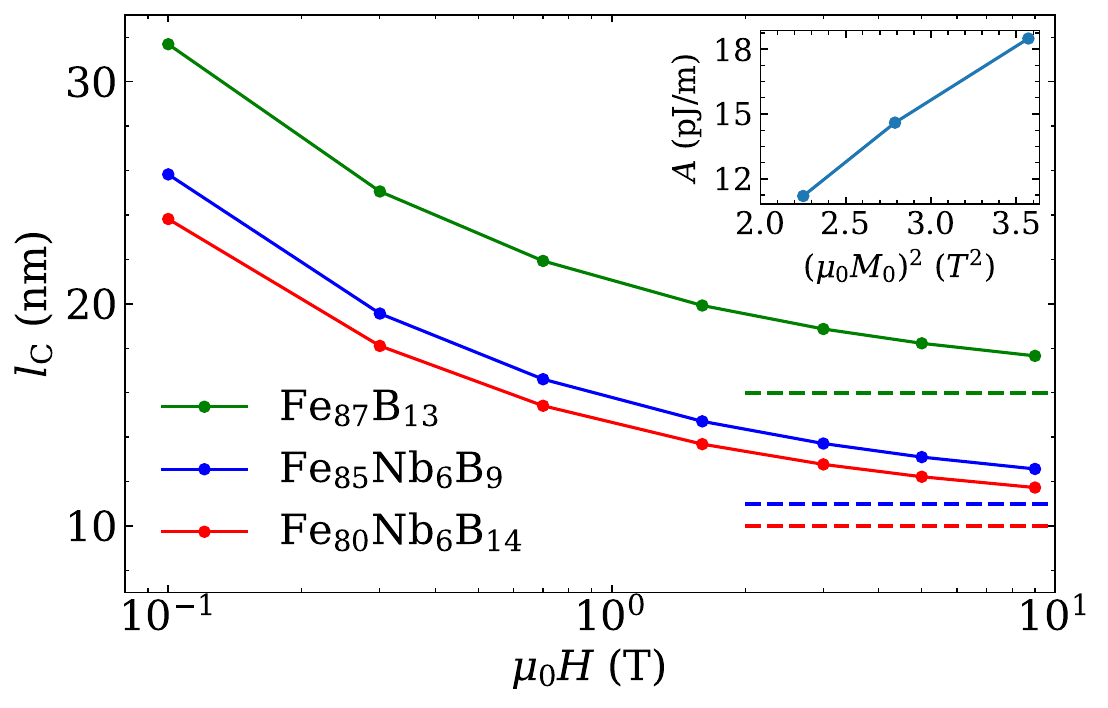}}
		\caption{Field variation of the correlation length $l_{\mathrm{C}}$ [Eq.~(\ref{lcdef})] (semi-log scale). The dashed lines indicate the average particle size $D$ (compare Table~\ref{tab:sample_char}). Inset:~Plot of $A$ versus $(\mu_0M_0)^2$.}
		\label{fig:micromag_length}
	\end{figure}

	The fit analysis provides the values of the exchange-stiffness constant $A$ as well as the average anisotropy field $\sqrt{\langle| \mathbf{H}_{\mathrm{p}}|^2 \rangle}$ and magnetostatic field $\sqrt{\langle| M_z |^2 \rangle}$ (see Table~\ref{tab:sans_1d_data}). We re-emphasize that the values for $\sqrt{\langle| \mathbf{H}_{\mathrm{p}}|^2 \rangle}$ and $\sqrt{\langle| M_z |^2 \rangle}$ are lower bounds, since the experimental data for $S_{\mathrm{H}}$ and $S_{\mathrm{M}}$ are only available within a finite range of momentum transfers between $q_{\mathrm{min}}$ and $q_{\mathrm{max}}$. It is also important to mention that these values represent effective values that are averaged over particle and matrix phases. The $A$~values are subsequently used to calculate the micromagnetic exchange lengths $l_{\mathrm{H}}$ [Eq.~(\ref{lhdef})], and the field dependence of the micromagnetic correlation length
	\begin{equation}
		\label{lcdef}
		l_{\mathrm{C}}(H_{\mathrm{i}}) = D + l_{\mathrm{H}}(H_{\mathrm{i}}) = D + \sqrt{\frac{2 A}{\mu_0 M_0 H_{\mathrm{i}}}} ,
	\end{equation}
	where $D$ is the average particle size, is depicted in Fig.~\ref{fig:micromag_length}. This length scale can be interpreted as the characteristic size regime over which perturbations in the spin structure are effectively mediated by the exchange interaction~\cite{michels03prl,bickapl2013,mettus2015}. It is seen that the $l_{\mathrm{C}}$ for all three samples exhibit a similar field variation with sizes ranging between about $10$$-$$35 \, \mathrm{nm}$. The behavior of the exchange constant appears to be in line with the expected $A \propto M_0^2$ scaling (see inset in Fig.~\ref{fig:micromag_length})~\cite{chikaani}.
	
	\begin{table}
		\small
		\centering
		\caption{Results of the micromagnetic SANS data analysis for the exchange-stiffness constant ($A$), mean-square anisotropy ($\sqrt{\langle| \mathbf{H}_{\mathrm{p}}|^2 \rangle}$), and mean-square magnetostatic field ($\sqrt{\langle| M_z |^2 \rangle}$).}
		\begin{tabular}{lccc}
			\hline
			\hline
			Alloy & $A$~(pJ/m) & $\mu_0 \sqrt{\langle| \mathbf{H}_{\mathrm{p}}|^2 \rangle}$~(mT) & $\mu_0 \sqrt{\langle| M_z |^2 \rangle}$~(mT)
			\tabularnewline[\doublerulesep]
			\hline 
			\noalign{\vskip\doublerulesep}
			Fe$_{87}$B$_{13}$ & $18.5 \pm 0.7$ & 83 & 207 
			\tabularnewline[\doublerulesep]
			Fe$_{85}$Nb$_6$B$_{9}$ & $14.6 \pm 0.2$ & 45 & 326
			\tabularnewline[\doublerulesep]
			Fe$_{80}$Nb$_6$B$_{14}$ & $10.9 \pm 0.2$ & 132 & 325
			\tabularnewline[\doublerulesep]
			\hline
		\end{tabular}
		\label{tab:sans_1d_data}		
	\end{table}
	
	Previously, we have utilized the present micromagnetic SANS data analysis to estimate the volume-averaged exchange-stiffness constants in a number of Fe-based nanocrystalline alloys. For Fe$_{89}$Zr$_7$B$_{3}$Cu, (Fe$_{0.985}$Co$_{0.015}$)$_{90}$Zr$_7$B$_{3}$~\cite{honecker2013analysis}, and (Fe$_{0.7}$Ni$_{0.3})$$_{86}$B$_{14}$~\cite{bersweiler2022unraveling}, we have found, respectively, $A = 3.1 \, \mathrm{pJ/m}$, $A = 4.7 \, \mathrm{pJ/m}$, and $A = 10.0 \, \mathrm{pJ/m}$. The present values in Table~\ref{tab:sans_1d_data} are somewhat larger than these, which might be surprising in view of the fact that a typical $A$~value for these types of alloys is often taken as $\sim$$10.0 \, \mathrm{pJ/m}$ (e.g., Ref.~\cite{herzer97}). However, it must be emphasized that (i)~the volume-averaged $A$~value may depend on the composition and on the microstructure (e.g., the heat treatment) and that (ii)~there exists a certain scatter in the values depending on the used experimental technique. To classify our experimental data we provide the following consideration. The value of $A$ at a given temperature can be computed based on the experimental value for the spin-wave stiffness constant $\mathcal{D}$, according to~\cite{coeybook2009}:
	\begin{eqnarray}
		\label{ADrelation}
		\mathcal{D} = \frac{2A g \mu_{\mathrm{B}}}{M_0} ,
	\end{eqnarray}
	where $g$ is the Land$\mathrm{\acute{e}}$ factor and $\mu_{\mathrm{B}}$ the Bohr magneton. The parameter $\mathcal{D}$ can e.g.\ be determined from inelastic neutron scattering, magnetization, or spin-wave resonance experiments. Using the room temperature value of $\mathcal{D} = 281 \, \mathrm{meV}${\AA}$^2$ for {\it single crystalline} Fe, obtained by triple-axis neutron spectroscopy~\cite{collins1969}, $g=2.10$ and $\mu_0 M_0 = 2.15 \, \mathrm{T}$~\cite{skomski2003nanomagnetics}, one obtains $A \cong 19.8 \, \mathrm{pJ/m}$. This value for pure Fe may be seen as an upper bound to our $A$~values obtained on Fe-based alloys.
	
	\begin{figure}[h]
		\centering
		\resizebox{0.90\columnwidth}{!}{\includegraphics{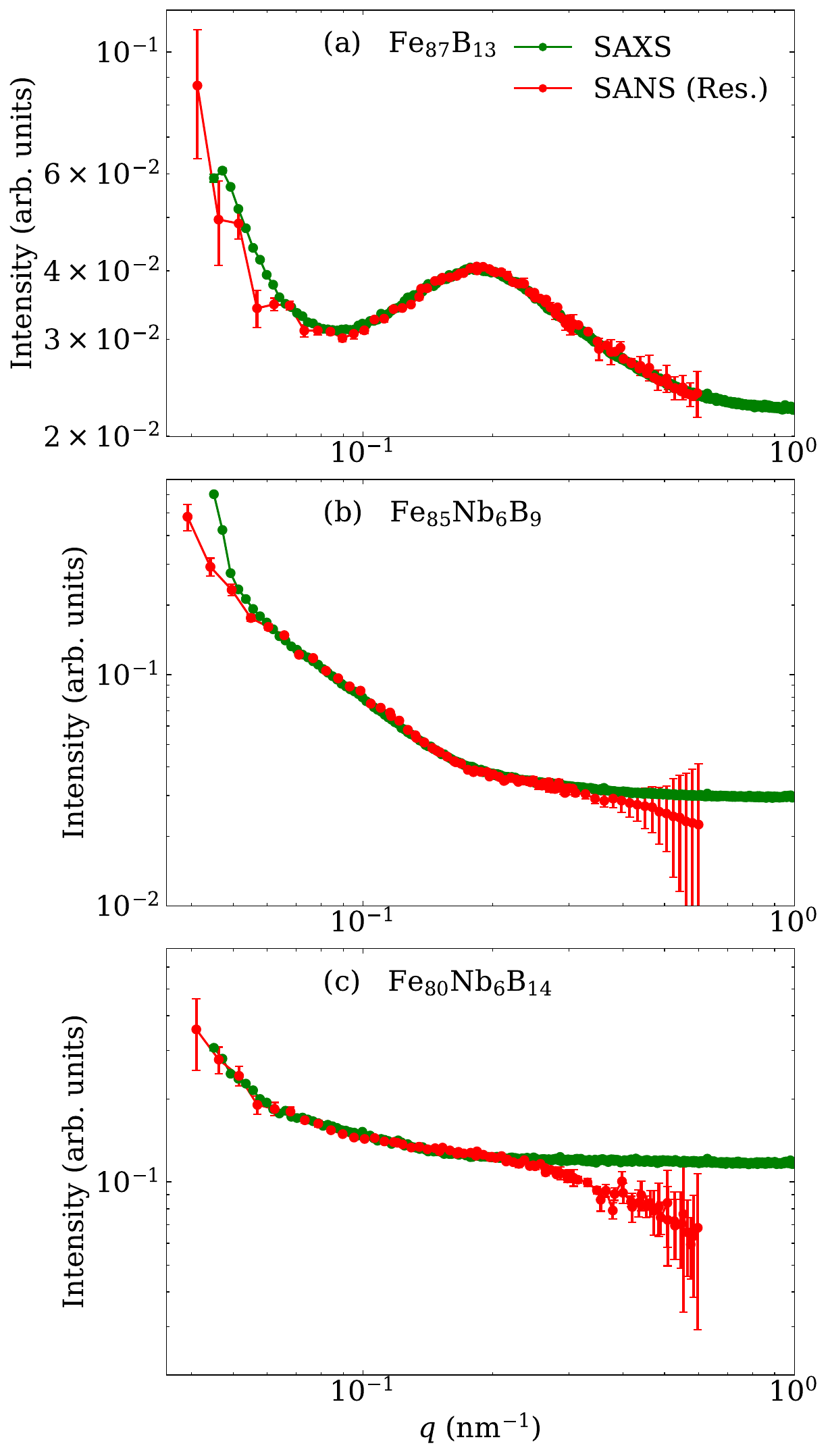}}
		\caption{(a)$-$(c) Comparison of the residual SANS cross sections $d \Sigma_{\mathrm{res}} / d \Omega$ of the three Nanoperm alloys (obtained from the micromagnetic neutron data analysis) with the respective SAXS signal (see insets) (log-log scale).}
		\label{fig:saxs_sans}
	\end{figure}
	
	The analysis of the azimuthally-averaged SANS data also provides the residual SANS cross section $d \Sigma_{\mathrm{res}} / d \Omega$, which represents the nuclear and longitudinal magnetic scattering at saturation. The $q$~dependence of the $d \Sigma_{\mathrm{res}} / d \Omega$ for all three Nanoperm samples is compared in Fig.~\ref{fig:saxs_sans} with the corresponding small-angle X-ray scattering (SAXS) cross sections. The shown $d \Sigma_{\mathrm{res}} / d \Omega$ have been scaled by constant factors to closely match the SAXS data. It can be seen that for Fe$_{87}$B$_{13}$ [Fig.~\ref{fig:saxs_sans}(a)] the $d \Sigma_{\mathrm{res}} / d \Omega$ matches very well with its SAXS cross section over the entire range of scattering vectors. The peak at $q_{\mathrm{max}} \cong 0.185 \, \mathrm{nm}^{-1}$ (corresponding to $2\pi/q_{\mathrm{max}} \cong 34 \, \mathrm{nm}$) is also observable in SAXS and reproduced by the micromagnetic SANS data analysis as well. The data in Fig.~\ref{fig:saxs_sans}(a) suggest that the nuclear and magnetic microstructure at saturation are ``congruent'', i.e., the structural features giving rise to the SAXS signal have essentially the same size, shape, and arrangement as the objects that are at the origin of $d \Sigma_{\mathrm{res}} / d \Omega$. For Fe$_{85}$Nb$_6$B$_{9}$ and Fe$_{80}$Nb$_6$B$_{14}$ [Fig.~\ref{fig:saxs_sans}(b) and (c)], both cross sections agree reasonably only at low and intermediate momentum transfers. This might be explained by the fact that the micromagnetic SANS data analysis becomes progressively more difficult at large $q$ (at least for these two samples), when the SANS cross section is independent of the field (compare the error bars in Fig.~\ref{fig:saxs_sans}).
	
	\section{Summary and Conclusion}
	\label{summary}
	
	We have conducted a combined SANS and SAXS study on a series of nanocrystalline two-phase Nanoperm alloys (Fe$_{87}$B$_{13}$, Fe$_{85}$Nb$_6$B$_{9}$, Fe$_{80}$Nb$_6$B$_{14}$). The SANS data can be very well described by micromagnetic SANS theory and yield values for the exchange-stiffness constants as well as estimates for the average anisotropy and magnetostatic fields. A distinct clover-leaf-shaped angular anisotropy in the magnetic SANS cross section strongly suggests that the magnetic scattering in these compounds predominantly stems from the magnetodipolar stray fields decorating the nanocrystals. This is in line with the observation that the average magnetostatic fields due to spatial variations in the longitudinal magnetization are much larger than the anisotropy fields. The obtained exchange-stiffness constants allowed us to draw conclusions on the field dependence of the micromagnetic exchange length, which is a measure for the size regime ($\sim$$10$$-$$35 \, \mathrm{nm}$) over which perturbations in the spin structure are effectively mediated by the exchange interaction. The nuclear and magnetic residual SANS cross sections, as obtained from the micromagnetic SANS data analysis, closely resemble the purely structural SAXS signal, providing further support for the validity of the micromagnetic SANS theory. Based on the neutron results we can state that the magnetic microstructure of the studied Nanoperm alloys is governed by static nanometer-scale long-wavelength magnetization fluctuations that have their origin in a highly nonuniform saturation magnetization profile $M_{\mathrm{s}}(\mathbf{r})$. From an application point of view, it might be beneficial to reduce the inhomogeneity in $M_{\mathrm{s}}(\mathbf{r})$, e.g., via the engineering of nanocrystallites that have a saturation magnetization equal or close to the matrix material, in this way reducing large jumps in $M_{\mathrm{s}}$.
	
	\section*{Acknowledgments}
	
	The authors acknowledge the Swiss spallation neutron source at the Paul Scherrer Institute for the provision of neutron beam time at the SANS-I instrument. M. P. Adams acknowledges financial support from the National Research Fund of Luxembourg (AFR Grant No.~15639149).
	
	\appendix
	\section*{Appendix: Overview of SANS results for $d \Sigma / d \Omega$ and $d \Sigma_{\mathrm{M}} / d \Omega$}
	\label{appendix}
	
	In this Appendix we display additional results for the two-dimensional total (nuclear and magnetic) SANS cross section $d \Sigma / d \Omega$ and for the purely magnetic SANS cross section $d \Sigma_{\mathrm{M}} / d \Omega$ of the Nanoperm alloys.
	
	\begin{figure*}
		\centering
		\resizebox{1.60\columnwidth}{!}{\includegraphics{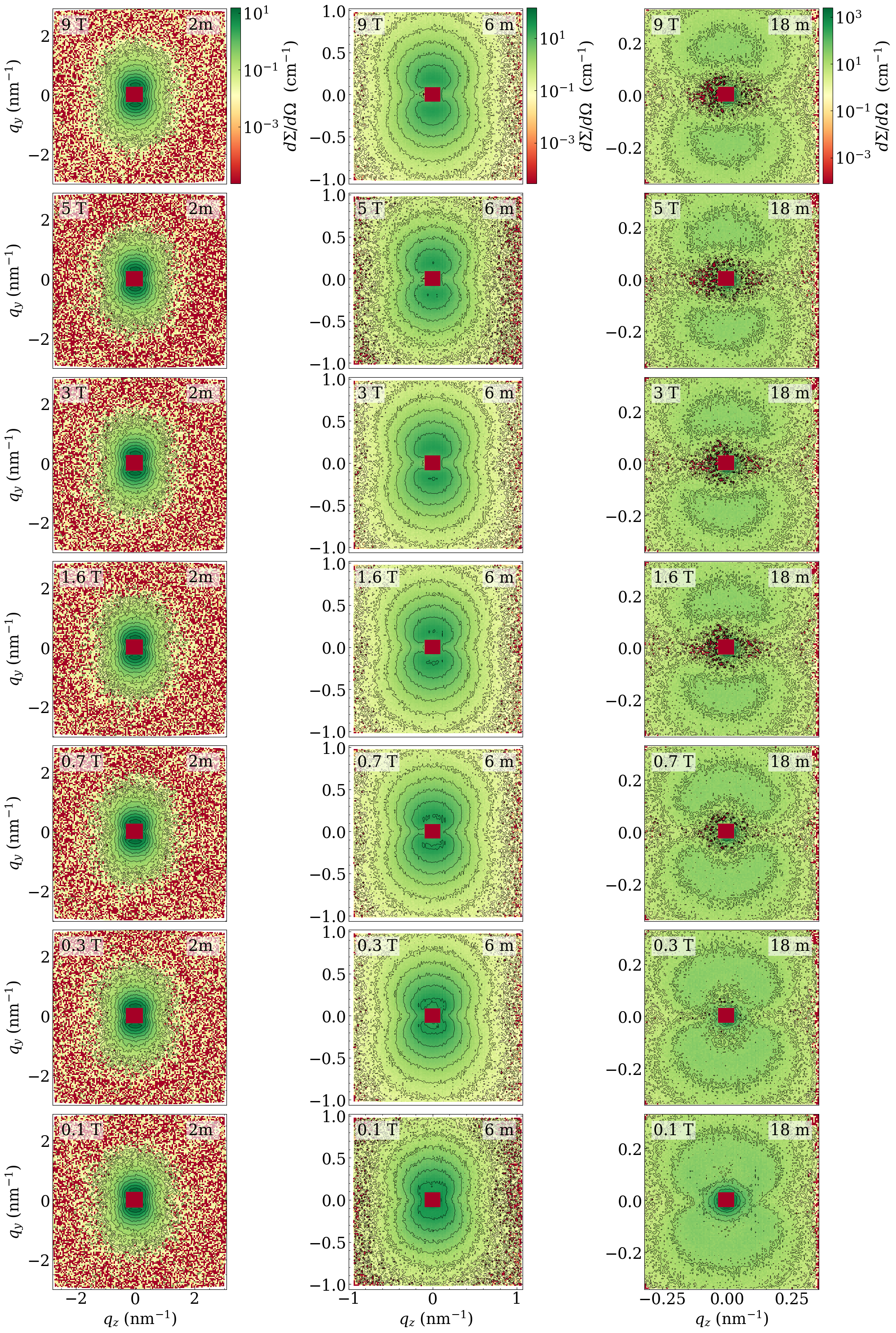}}
		\caption{Total nuclear and magnetic SANS cross section $d \Sigma / d \Omega$ of Fe$_{87}$B$_{13}$ alloy at selected applied magnetic fields (see insets) (logarithmic color scale). The sample-to-detector distance varies from the left to the right column ($2 \, \mathrm{m}$, $6 \, \mathrm{m}$, $18 \, \mathrm{m}$).}
		\label{fig:SANS_2D_K1_paper}
	\end{figure*}
	
	\begin{figure*}
		\centering
		\resizebox{1.60\columnwidth}{!}{\includegraphics{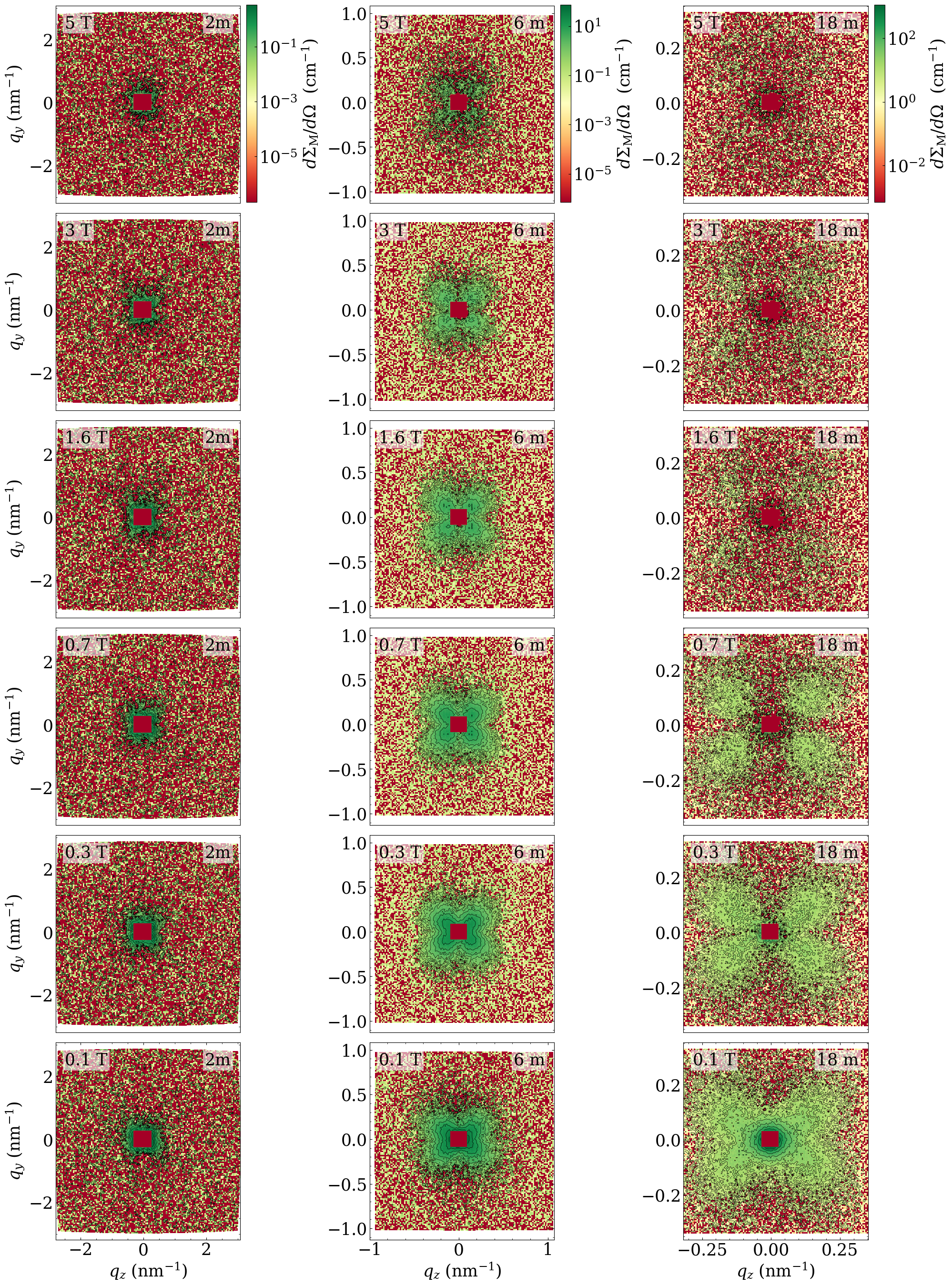}}
		\caption{Magnetic SANS cross section $d \Sigma_{\mathrm{M}} / d \Omega$ of Fe$_{87}$B$_{13}$ alloy at selected applied magnetic fields (see insets) (logarithmic color scale). The nuclear and magnetic scattering at $9 \, \mathrm{T}$ (saturation) has been subtracted from each data set.}
		\label{fig:SANS_2D_mag_K1_paper}
	\end{figure*}
	
	\begin{figure*}
		\centering
		\resizebox{1.60\columnwidth}{!}{\includegraphics{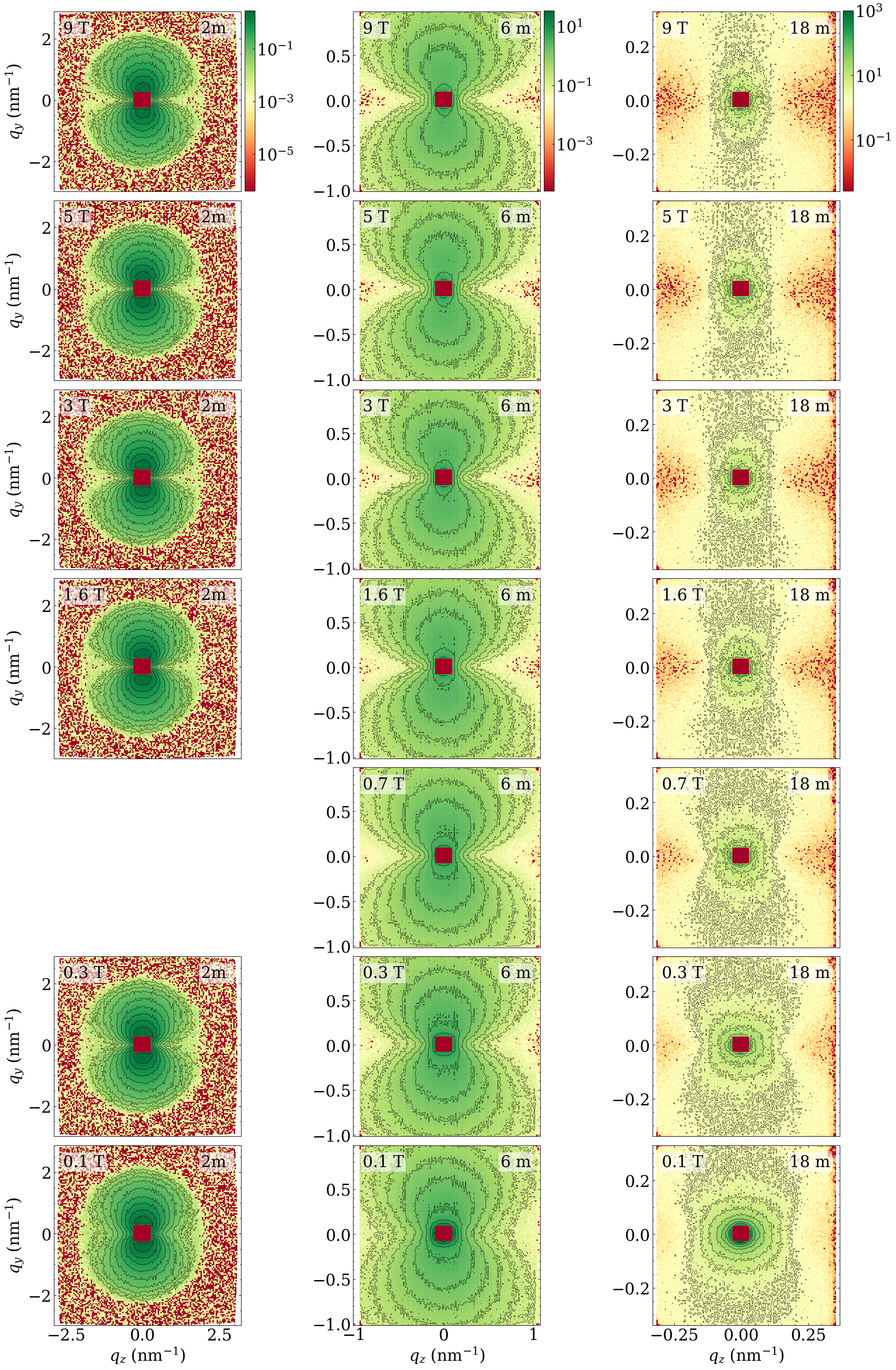}}
		\caption{Total nuclear and magnetic SANS cross section $d \Sigma / d \Omega$ of Fe$_{85}$Nb$_6$B$_{9}$ alloy at selected applied magnetic fields (see insets) (logarithmic color scale).}
		\label{fig:SANS_2D_K2_paper}
	\end{figure*}
	
	\begin{figure*}
		\centering
		\resizebox{1.60\columnwidth}{!}{\includegraphics{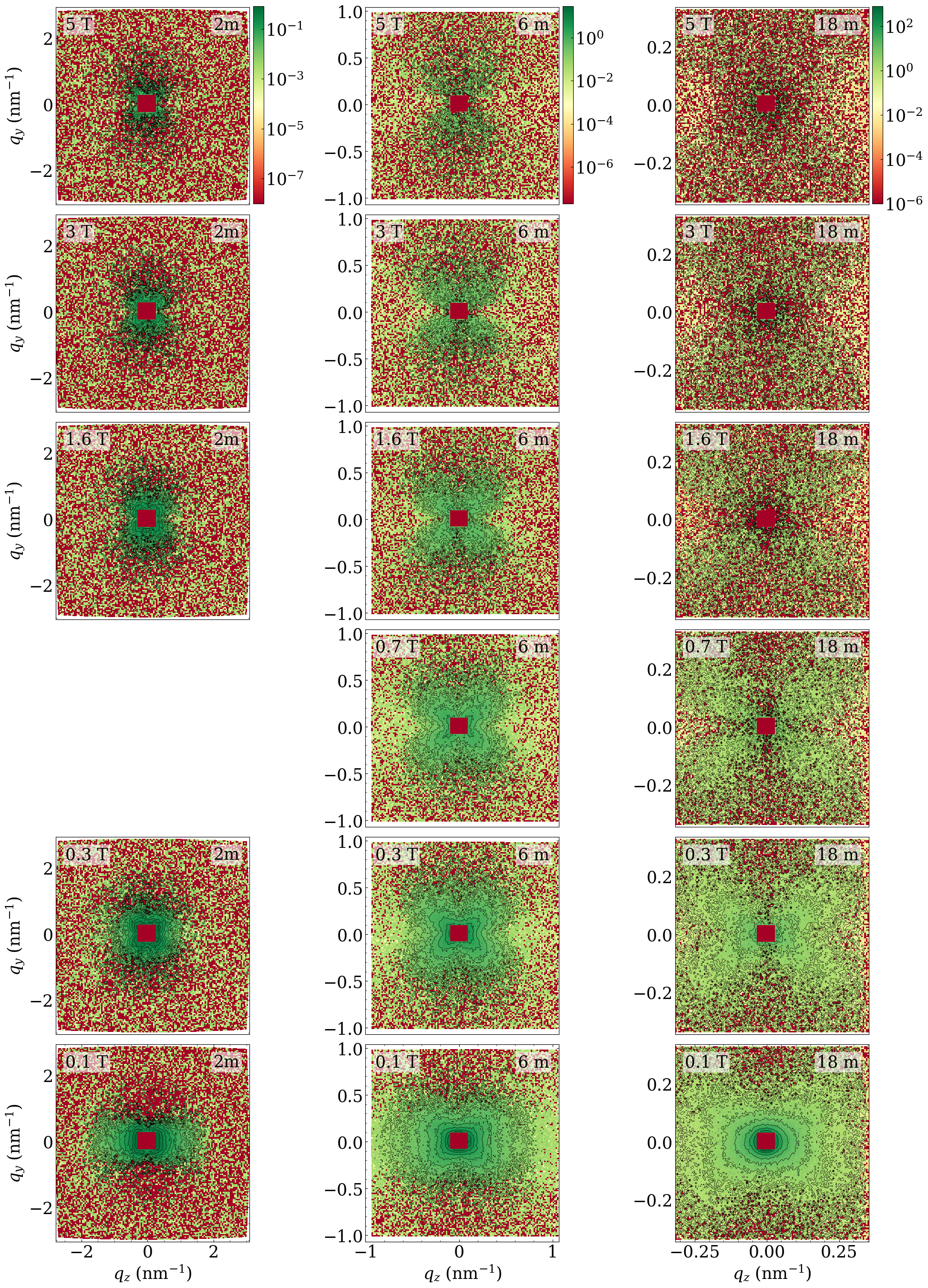}}
		\caption{Magnetic SANS cross section $d \Sigma_{\mathrm{M}} / d \Omega$ of Fe$_{85}$Nb$_6$B$_{9}$ alloy at selected applied magnetic fields (see insets) (logarithmic color scale). The nuclear and magnetic scattering at $9 \, \mathrm{T}$ (saturation) has been subtracted from each data set.}
		\label{fig:SANS_2D_mag_K2_paper}
	\end{figure*}
	
	\begin{figure*}
		\centering
		\resizebox{1.60\columnwidth}{!}{\includegraphics{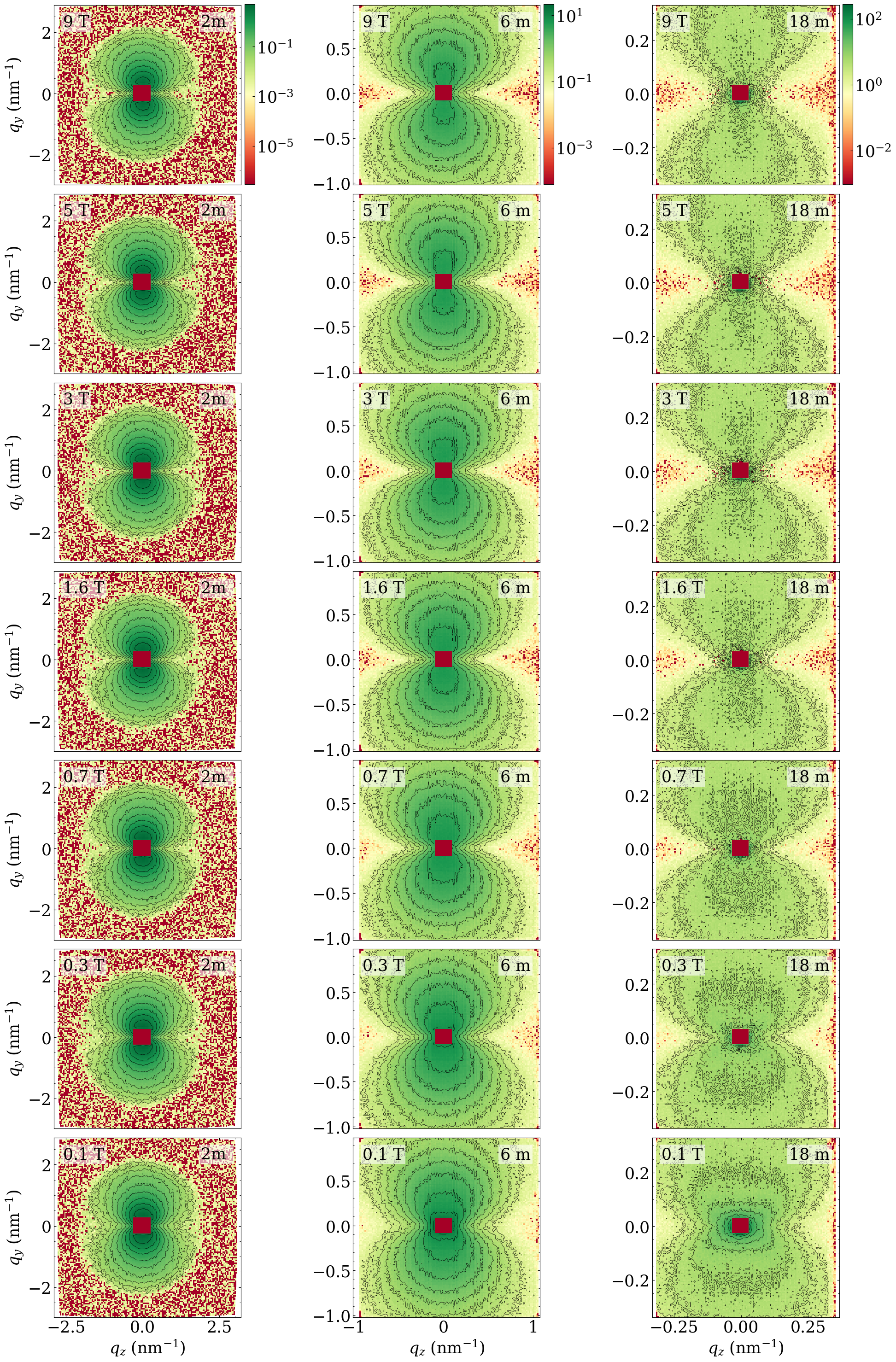}}
		\caption{Total nuclear and magnetic SANS cross section $d \Sigma / d \Omega$ of Fe$_{80}$Nb$_6$B$_{14}$ alloy at selected applied magnetic fields (see insets) (logarithmic color scale).}
		\label{fig:SANS_2D_K3_paper}
	\end{figure*}
	
	\begin{figure*}
		\centering
		\resizebox{1.60\columnwidth}{!}{\includegraphics{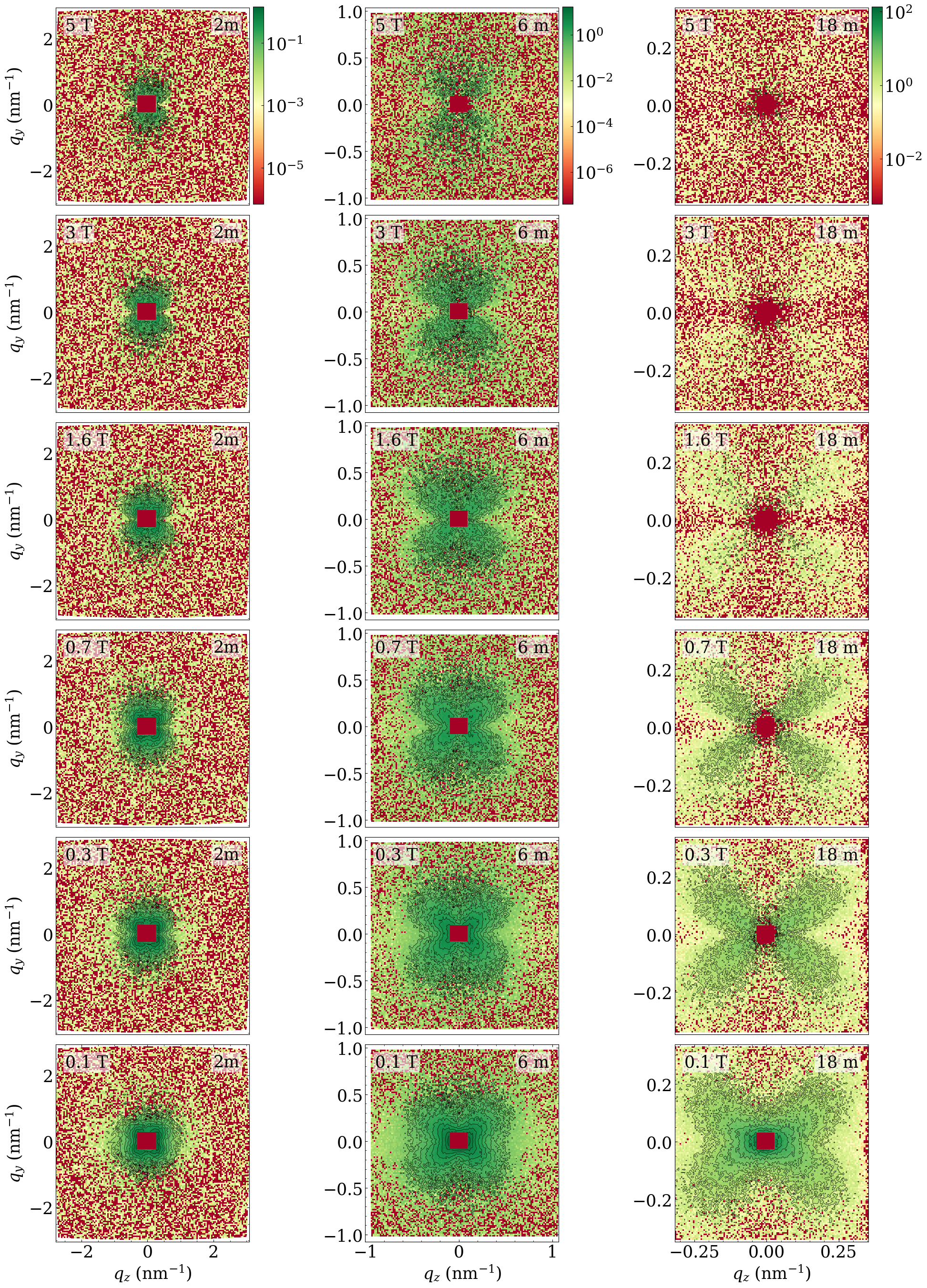}}
		\caption{Magnetic SANS cross section $d \Sigma_{\mathrm{M}} / d \Omega$ of Fe$_{80}$Nb$_6$B$_{14}$ alloy at selected applied magnetic fields (see insets) (logarithmic color scale). The nuclear and magnetic scattering at $9 \, \mathrm{T}$ (saturation) has been subtracted from each data set.}
		\label{fig:SANS_2D_mag_K3_paper}
	\end{figure*}
	
	\bibliography{Nanoperm_K123_2023}
	
\end{document}